\documentclass[showpacs,pra,aps,superscriptaddress,twocolumn,floatfix]{revtex4}
\usepackage{amsmath}
\usepackage{amssymb,mathrsfs}
\usepackage{graphicx}% Include figure files

\newcommand{\al}{\alpha}
\newcommand{\be}{\beta}
\newcommand{\la}{\lambda}
\newcommand{\ga}{\gamma}
\newcommand{\vphi}{\varphi}
\newcommand{\prt}{\partial}

\begin{document}

\title{Nonlinear waves in two-component Bose-Einstein condensates: Manakov
system and Kowalevski equations}

\author{A. M. Kamchatnov}
\affiliation{Institute of Spectroscopy,
  Russian Academy of Sciences, Troitsk, Moscow, 142190, Russia}
\author{V. V. Sokolov}
\affiliation{L. D. Landau Institute for Theoretical Physics, Russian Academy of Sciences,
Moscow Region, 142432, Russia}

\date{\today}

\begin{abstract}
  Traveling waves in two-component Bose-Einstein condensates whose dynamics is described by the Manakov
  limit of the Gross-Pitaevskii equations are considered in general situation with relative motion of
  the components when their chemical potentials are not equal to each other. It is shown that in this case
  the solution is reduced to the form known in the theory of motion of S.~Kowalevski top.
  Typical situations are illustrated by the particular cases when the general solution can be represented in
  terms of elliptic functions and their limits. Depending on the parameters of the wave, both density
  waves (with in-phase motions of the components) and polarization waves (with counter-phase
  their motions) are considered.
\end{abstract}


\maketitle

\section{Introduction}

Realization of Bose-Einstein condensates (BECs) of atoms which can occupy several quantum states at extremely low temperatures
has drawn interest to nonlinear dynamics of such multi-component systems (see, e.g., the review article \cite{ar-13}).
In particular, if the components can mix, then in such a condensate two types of linear waves can propagate---usual
sound waves (that is, density waves) with in-phase oscillations of the components and so-called {\it polarization waves}
with counter-phase oscillations. Correspondingly, generally speaking, there exist two Mach cones and two channels of
Cherenkov radiation what leads to considerable changes in the character of excitations in the system compared with one-component
situation.
The same holds true for the nonlinear excitations---solitons and breathers. For example, oblique solitons can be
generated by the flow of two-component BEC past a non-polarized obstacle which repels both components \cite{gladush-09}
and oblique breathers are generated in the case of polarized obstacles which repel one component and attract the other
one \cite{kk-13}.

The dynamics of two-component BECs becomes even richer if one component moves with respect to another.
As was found experimentally in \cite{hamner-11,hoefer-11,hamner-13}, the relative motion of components leads to
generation of nonlinear periodic waves of polarization. In these experiments the components correspond to the
quantum states $|1,-1\rangle$ and $|2,-2\rangle$ of the hyperfine structure of $^{87}\mathrm{Rb}$ atoms with very close
values of their inter-atomic scattering lengths. Hence their quasi-one-dimensional dynamics in elongated
cigar-shaped traps can be described with high accuracy by the Gross-Pitaevskii equations of the Manakov type \cite{manakov}.
In the standard non-dimensional units these equations can be written in the form
\begin{equation}\label{eq1}
    i\psi_{k,t}+\frac12\psi_{k,xx}-(|\psi_1|^2+|\psi_2|^2)\psi_k=0,\qquad k=1,2,
\end{equation}
where $\psi_k$ denotes the wave function of the $k$th component, $x$ is the coordinate along the trap, and $t$ is the time
variable. Such multi-component (vector) nonlinear Schr\"odinger equations
appeared first in nonlinear optics \cite{bz-70} and they have been studied intensely in this physical
context, where  it is natural to suppose that the wave numbers (which are analogues of velocities of
the BEC components) of both components are equal to each other (see, e.g., \cite{ka-2003}), but the interaction constants
are different. Thus, so far the situation with equal nonlinearity constants and non-vanishing relative velocity of the
components was studied very little. An important particular case of such a type of solutions is so-called {\it dark-bright soliton}
with vanishing of one of the background densities far enough from the soliton location. This means that the component with
non-vanishing background density forms a ``trap'' for another component localized inside such a trap (see, e.g.,
\cite{sk-1997,{smkj-10}}). Although this solution of the Manakov system describes an important type of nonlinear
excitations in two-component BECs, it cannot explain dense lattices of dark-bright solitons observed in
recent experiments \cite{hamner-11,hoefer-11,hamner-13}. An attempt of such an explanation was done in
Refs.~\cite{wright-13,kamch-13,kamch-14,kamch-14b} where particular cases of nonlinear waves with relative motions of
the components were studied. Indeed, solutions in the form of counter-phase oscillations (polarization waves)
were found in these papers, however, they were limited to BECs with equal chemical potentials and
this condition is quite restrictive for adequate description of experimental observations.

In this paper we shall consider the general situation with non-equal chemical potentials
for the case of one-phase traveling waves. It will be shown that in this case the Manakov system can be reduced
to the equations studied first by Sophie Kowalevski in her theory of rotation of the so-called {\it Kowalewski top}
\cite{kow-1889,golubev}. (Similar reduction was performed for the Manakov system with attractive  (focusing)
nonlinear interaction and without relative motion of the components in Refs.~\cite{phf-98,cezk-2000,eek-2000}.)

The physical conditions that the densities of the components must be
positive and non-singular impose heavy restrictions on the admissible solutions of the Kowalevski equations.
The typical situation will be illustrated by several particular cases. In particular, it will be shown
that the solutions studied previously in \cite{wright-13,kamch-13,kamch-14,kamch-14b} can be obtained as
a special limiting case of the general solution of the Kowalevski equations.
Another particular case of the dark-bright solitons is also obtained as a result of simple degeneration of the
Kowalevski equations.
The so-called {\it Appelrot-Delone class} of solutions of the Kowalevski
equations, studied previously in the context of rotations of the Kowalevski top, leads now in the
context of two-component condensate flows to the nonlinear density waves with very
specific dispersion law conditioned by the relative motion of the components. In the soliton limit this
periodic solution reduces to the dark-dark solitons which generalize the known Manakov soliton solution
to the situation with relative motion of the components.
The general nonlinear wave in the two-component BEC is illustrated by
the so-called Legendre-Jacobi case when the hyperelliptic integrals are reduced to the elliptic ones.
The physical implications of the found solutions are
discussed in the conclusion of the paper.

\section{Equations of motion, their integrals and general solution}

It is convenient to transform the Manakov system (\ref{eq1}) to the hydrodynamic-like form by means of the
Madelung transformation
\begin{equation}\label{eq2}
    \psi_k=\sqrt{\rho_k}\exp\left(i\int^xu_kdx-i\mu_kt\right),
\end{equation}
where $\mu_{1,2}$ are constants.  In a standard way we arrive at the system
\begin{equation}\label{eq3}
    \begin{split}
    &\rho_{k,t}+(\rho_ku_k)_x=0,\\
    &u_{k,t}+\left(\frac12u_k^2+\rho_1+\rho_2+\frac{\rho_{k,x}^2}{8\rho_k^2}-\frac{\rho_{k,xx}}{4\rho_k}\right)_x=0
    \end{split}
\end{equation}
with real variables. Here $\rho_{1,2}$ denote the densities of the components, $u_{1,2}$ denote their flow
velocities, and $\mu_{1,2}$ are their chemical potentials.

The complex substitution $v_k=u_k-i\rho_{k,x}/(2\rho_k)$ casts the system (\ref{eq3}) to a mathematically simpler form
\begin{equation}\label{eq4}
    \begin{split}
    &\rho_{k,t}+\left(\rho_kv_k+\frac{i}2\rho_{k,x}\right)_x=0,\\
    &v_{k,t}+\left(\frac12v_k^2+\rho_1+\rho_2-\frac{i}2v_{k,x}\right)_x=0.
    \end{split}
\end{equation}
A traveling wave is described by one-phase solution of Eqs.~(\ref{eq4}) with all the variables depending on $\xi=x-Vt$ only,
\begin{equation}\label{eq5}
    \rho_k=\rho_k(\xi),\quad v_k=v_k(\xi),\quad \xi=x-Vt,
\end{equation}
where $V$ is a constant velocity of the wave. Let us introduce the imaginary
``time'' variable $\tau=-2i\xi$. Then after obvious integrations we get
\begin{equation}\label{eq6}
    \frac{d\rho_k}{d\tau}=\al_k-\rho_kw_k,\quad \frac{dw_k}{d\tau}=\frac12w_k^2+\rho_1+\rho_2-\be_k,
\end{equation}
where $\alpha_k$, $\be_k$ are the integration constants  and $w_k=v_k-V$.

It is worth noticing that integration of the first pair of Eqs.~(\ref{eq3})
under {\it ansatz} (\ref{eq5}) gives expressions for $u_k$ in terms of $\rho_k$,
\begin{equation}\label{eq6a}
    u_k(\xi)=V+\frac{\al_k}{\rho_k(\xi)},\quad k=1,2.
\end{equation}
Similar integration of the second pair of Eqs.~(\ref{eq3}) followed by exclusion of $u_k$ with the use
of Eqs.~(\ref{eq6a}) yields
\begin{equation}\label{eq6b}
    \frac{\rho_{k,\xi}^2}{8\rho_k^2}-\frac{\rho_{k,\xi\xi}}{4\rho_k}+\frac{\al_k^2}{2\rho_k^2}+\rho_1+\rho_2=\be_k.
\end{equation}
As follows from these relations, the constants $\al_k$, $\be_k,$ $k=1,\,2,$ are real. In case of a uniform flow
with constant $\rho_k=\rho_{k0}$ and $u_k=u_{k0}$ we have $\be_k=\rho_{10}+\rho_{20}+\alpha_k^2/(2\rho_{k0}^2)$.
Hence, the parameters $\be_k$ are related with the chemical potentials $\mu_k=u_k^2/2+\rho_{10}+\rho_{20}$ by the
formulae
\begin{equation}\label{eq6c}
    \mu_k=\frac12V^2+\frac{V\alpha_k}{\rho_{k0}}+\be_k,\quad k=1,2.
\end{equation}
The solution studied in Refs.~\cite{wright-13,kamch-13,kamch-14,kamch-14b} were limited to the case $\be_1=\be_2$
what meant that in the uniform case the chemical potentials are equal to each other: $\mu_1=\mu_2$. Here we will
consider the general case including situations with $\be_1\neq\be_2$.

Our approach is based on the fact that the system (\ref{eq6}) is Hamiltonian with the Hamiltonian
\begin{equation}\label{eq7}
\begin{split}
    H=&-\frac12(\rho_1w_1^2+\rho_2w_2^2)-\frac12(\rho_1+\rho_2)^2\\
    &+\al_1w_1+\al_2w_2+\be_1\rho_1+\be_2\rho_2
    \end{split}
\end{equation}
and Poisson brackets
\begin{equation}\label{eq8}
    \left\{\rho_i,\rho_j\right\}=\left\{w_i,w_j\right\}=0,\quad \left\{w_i,\rho_j\right\}=\delta_{ij}.
\end{equation}
The corresponding equations of motion
\begin{equation}\label{eq9}
    \frac{d\rho_k}{d\tau}=\frac{\prt H}{\prt w_k},\quad \frac{dw_k}{d\tau}=-\frac{\prt H}{\prt \rho_k}.
\end{equation}
possess the integral of energy
\begin{equation}\label{eq10}
    H(\rho_k,w_k)=h=\mathrm{const}.
\end{equation}
For complete integrability of this system with two degrees of freedom we need, according to the Liouville-Arnold
theorem (see, e.g., \cite{arnold}), one more integral.
One can check that there is such an integral quadratic in momenta $w_k$:
\begin{equation}\label{eq11}
    \begin{split}
    K=&-\rho_1\rho_2(w_1-w_2)^2+2(w_1-w_2)(\al_1\rho_2-\al_2\rho_1)\\
    &-(\be_1-\be_2)[\rho_1w_1^2-\rho_2w_2^2+\rho_1^2-\rho_2^2\\
    &-2(\al_1w_1-\al_2w_2)-2(\be_1\rho_1-\be_2\rho_2)].
    \end{split}
\end{equation}
Thus, integration of the system (\ref{eq9}) can be reduced to
quadratures.

If $\be_1\neq\be_2$, then we can make a canonical transformation
\begin{equation}\label{eq17}
\begin{split}
    &\rho_1=\frac{(q_1+\be)(q_2+\be)}{2\be},\quad \rho_2=-\frac{(q_1-\be)(q_2-\be)}{2\be},\\
    &w_1=\frac{(q_1-\be)p_1-(q_2-\be)p_2}{q_1-q_2}\\
    &w_2=\frac{(q_1+\be)p_1-(q_2+\be)p_2}{q_1-q_2},
    \end{split}
\end{equation}
where we have denoted $\be\equiv\be_1-\be_2$ (the limit $\beta\to0$ will be considered below).
As we shall see, the dynamics is separable in these new variables $q_i,\,p_i$, $i=1,\,2$. 
The Poisson brackets preserve their canonical form
\begin{equation}\label{eq18}
    \{q_i,q_j\}=\{p_i,p_j\}=0,\quad \{p_i,q_j\}=\delta_{ij},
\end{equation}
and the  Hamiltonian becomes
\begin{equation}\label{eq19}
    \begin{split}
    &H=\frac{(q_1^2-\be^2)p_1^2-2[(\al_1+\al_2)q_1-(\al_1-\al_2)\be]p_1}{2(q_2-q_1)}\\
    &+\frac{(q_2^2-\be^2)p_2^2-2[(\al_1+\al_2)q_2-(\al_1-\al_2)\be]p_2}{2(q_1-q_2)}\\
    &-\frac12[q_1^2+q_2^2+q_1q_2-(\be_1+\be_2)(q_1+q_2)-\be^2].
    \end{split}
\end{equation}
The equations of motions are given by
\begin{equation}\label{eq20}
    \begin{split}
    \frac{dq_1}{d\tau}&=\frac{\prt H}{\prt p_1}\\
    &=\frac{(q_1^2-\be^2)p_1+(\al_1-\al_2)\be-(\al_1+\al_2)q_1}{q_1-q_2},\\
    \frac{dp_1}{d\tau}&=-\frac{\prt H}{\prt q_1}\\
    &=\frac{(p_1-p_2)[(\al_1-\al_2)\be-(\al_1+\al_2)q_2]}{(q_1-q_2)^2}\\
    &-\frac{(q_1^2+\be^2)p_1^2-(q_2^2-\be^2)p_2^2+2q_1q_2p_1^2}{2(q_1-q_2)^2}\\
    &+\frac12(\be_1+\be_2-2q_1-q_2),
    \end{split}
\end{equation}
and similar equations can be written for $q_2$ and $p_2$. They have two integrals of motion---the
energy $H(q_1,p_1,q_2,p_2)=h=\mathrm{const}$ and
\begin{equation}\label{eq21}
\begin{split}
    &K(q_1,p_1,q_2,p_2)=\big\{2(\al_1+\al_2)(p_1-p_2)q_1q_2\\
    &-(p_1^2q_1-p_2^2q_2)q_1q_2+\be^2(p_1^2q_2-p_2^2q_1)\\
    &-2\be(\al_1-\al_2)(p_1q_2-p_2q_1)\big\}/(q_1-q_2)\\
    &+(\be_1+\be_2-q_1-q_2)q_1q_2=k=\mathrm{const}.
    \end{split}
\end{equation}

As was mentioned above, integration of Hamiltonian system with two degrees of
freedom and two integrals of motion can be reduced to quadratures. Actual integration can be
performed in our case as follows. Eliminating
variables $q_2,\,p_2$ from the integrals $H(q,p)=h$ and $K(q,p)=k,$ we obtain
\begin{equation}\label{eq22}
\begin{split}
    &\Phi =(q_1^2-\be^2)(\be_1+\be_2-q_1-p_1^2)\\
    &+2p_1[(\al_1+\al_2)q_1-\be(\al_1-\al_2)]-2hq_1+k=0,
    \end{split}
\end{equation}
what demonstrates the mentioned above separation of variables.
Taking into account Eqs.~(\ref{eq20}), we get
\begin{equation}\label{eq23}
\begin{split}
    \frac{\prt\Phi}{\prt p_1}&=2(q_1^2-\be^2)(\be_1+\be_2-q_1-p_1^2)\\
    &-2[\be(\al_1-\al_2)-(\al_1+\al_2)q_1]\\
    &=-2(q_1-q_2)\frac{dq_1}{d\tau}.
    \end{split}
\end{equation}
 We solve equation (\ref{eq22}) with respect to
$p_1$ and substitute the result into (\ref{eq23}). After this and similar manipulations with the variables
$q_2,\,p_2$ we obtain the system
\begin{equation}\label{eq24}
    \begin{split}
    \pm\sqrt{-\mathcal{R}(q_1)}&=-(q_1-q_2)\frac{dq_1}{d\tau},\\
    \pm\sqrt{-\mathcal{R}(q_2)}&=(q_1-q_2)\frac{dq_2}{d\tau},
    \end{split}
\end{equation}
where
\begin{equation}\label{eq25}
\begin{split}
    \mathcal{R}(q)=&q^5-(\be_1+\be_2)q^4-2(\be^2-h)q^3\\
    &-[(\al_1+\al_2)^2-2(\be_1+\be_2)\be^2+k]q^2\\
    &+\be[\be^3-2h\be+2(\al_1^2-\al_2^2)]q\\
    &-\be^2[(\be_1+\be_2)\be^2-k+(\al_1-\al_2)^2]
    \end{split}
\end{equation}
is a 5th degree polynomial with respect to $q$. Then after simple manipulations we arrive at the system
\begin{equation}\label{eq26}
\begin{split}
    &\frac{dq_1}{\sqrt{\mathcal{R}(q_1)}}+\frac{dq_2}{\sqrt{\mathcal{R}(q_2)}}=0,\\
    &\frac{q_1dq_1}{\sqrt{\mathcal{R}(q_1)}}+\frac{q_2dq_2}{\sqrt{\mathcal{R}(q_2)}}=\pm 2\,{d\xi},
    \end{split}
\end{equation}
where we have returned to the real variable $\xi=i\tau/2$.
Sometimes it is convenient to rewrite this system in the  Kowalevski form
\begin{equation}\label{eq26a}
    \frac{dq_1}{d\xi}=\frac{2\sqrt{\mathcal{R}(q_1)}}{q_1-q_2},\quad
    \frac{dq_2}{d\xi}=-\frac{2\sqrt{\mathcal{R}(q_2)}}{q_1-q_2}.
\end{equation}

The systems (\ref{eq26}) or (\ref{eq26a}) can be solved formally in terms of Riemann $\theta$-functions
(more precisely, in terms of G\"opel and Rosenhein hyperelliptic functions; modern exposition of this method
can be found, e.g., in \cite{bbeim}) but such a form of the general solution is
mathematically involved and hardly can produce essential understanding of physical behavior
of waves in a two-component BEC. Therefore we shall confine ourselves here to the most
important particular solutions which provide useful information about such typical
nonlinear excitations in BEC as periodic waves and solitons.

\section{Nonlinear waves in a two-component BEC}

The physical variables $\rho_1$ and $\rho_2$ (i.e. densities of BEC components) must be
positive and this condition imposes important restrictions on the variables $q_1$ and $q_2$
which obey the systems (\ref{eq26}) or (\ref{eq26a}). Supposing for definiteness that $\be>0$,
it is easy to find that $q_1$ and $q_2$ can vary in the intervals (see Fig.~1)
\begin{equation}\label{eq27}
\begin{split}
    -\be&\leq q_1\leq\be,\quad q_2\geq\be,\\
    \text{or}\quad q_1&\geq\be,\quad-\be\leq q_2\leq\be.
    \end{split}
\end{equation}
\begin{figure}[ht]
\begin{center}
\includegraphics[width=6.5cm]{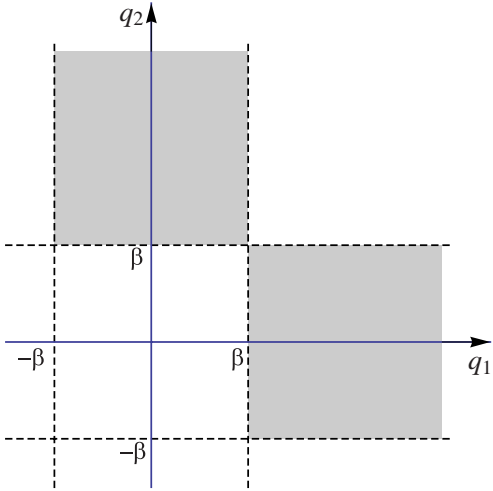}\\
\caption{Regions of variations of the parameters $q_1$ and $q_2$ (see Eq.~(\ref{eq27}))
corresponding to the conditions of positivity of the densities $\rho_1$ and $\rho_2$
defined by Eqs.~(\ref{eq17}).
}
\end{center}
\label{fig1}
\end{figure}

Formal integration of the system (\ref{eq26}) yields
\begin{equation}\label{eq27a}
\begin{split}
    &\int_{q_{10}}^{q}\frac{dq}{\sqrt{\mathcal{R}(q)}}+\int_{q_{20}}^{q_2}\frac{dq}{\sqrt{\mathcal{R}(q)}}=0,\\
    &\int_{q_{10}}^{q}\frac{qdq}{\sqrt{\mathcal{R}(q)}}+\int_{q_{20}}^{q_2}\frac{qdq}{\sqrt{\mathcal{R}(q)}}=\pm2{\xi},
     \end{split}
\end{equation}
where $q_{10}$ and $q_{20}$ are integration constants equal to the values of $q_1$ and $q_2$ at $\xi=0$,
respectively. Every solution of the system (\ref{eq26}) is parameterized by five zeroes $\nu_i$, $i=1,\ldots,5$, of the polynomial
\begin{equation}\label{eq27c}
    \mathcal{R}(q)=\prod_{i=1}^5(q-\nu_i)
\end{equation}
and, depending on their values, we obtain different classes of solutions.
Since the solution is symmetric with respect to transposition of $q_1$ and $q_2$,
for definiteness we assume that they change in the intervals (the zeroes $\nu_i$ are
numerated in the order of their values)
\begin{equation}\label{eq27b}
    \nu_1\leq q_1\leq\nu_2 \quad\text{and}\quad \nu_3\leq q_2\leq\nu_4
\end{equation}
where the polynomial $\mathcal{R}(q)$ is positive.

\subsection{Limit  $\be\to0$}

If $\be\equiv\be_1-\be_2\to0$, then  we must have $\nu_1\to0$ and $\nu_2\to0$ to satisfy the
conditions (\ref{eq27}). At the same time, to get in this singular limit from Eqs.~(\ref{eq17}) finite
values of $\rho_1$ and $\rho_2$, we have to define new variables and parameters as
\begin{equation}\label{eq-l1}
    q_1=\be\widetilde{q}_1,\quad \nu_1=\be\widetilde{\nu}_1,\quad \nu_2=\be\widetilde{\nu}_2
\end{equation}
so that Eqs.~(\ref{eq17}) reduce to
\begin{equation}\label{eq-l2}
    \rho_1=\frac12q_2(1+\widetilde{q}_1),\quad \rho_2=\frac12q_2(1-\widetilde{q}_1).
\end{equation}
The Kowalevski equations (\ref{eq26a}) are then transformed to
\begin{equation}\label{eq-l3}
    \begin{split}
    \frac{d\widetilde{q}_1}{d\xi}&=\frac{2\sqrt{\nu_3\nu_4\nu_5}}{q_2}\sqrt{(\widetilde{q}_1-\widetilde{\nu}_1)
    (\widetilde{\nu}_2-\widetilde{q}_2)},\\
    \frac{dq_2}{d\xi}&=-2\sqrt{(q_2-\nu_3)(\nu_4-q_2)(\nu_5-q_2)}.
    \end{split}
\end{equation}
We notice that the expression $\rho\equiv\rho_1+\rho_2=q_1+q_2$ reduces  to $\rho=q_2$ in this limit. Introducing
also $\widetilde{q}_1=\cos\theta$, $\widetilde{\nu}_1=\cos\theta_1$, $\widetilde{\nu}_2=\cos\theta_2$, $\nu_3=\rho_1$,
$\nu_4=\rho_2$, $\nu_5=\rho_3$,
we arrive at the equations
\begin{equation}\label{eq-l4}
\begin{split}
    \frac{d\cos\theta}{d\xi}&=-\frac{2\sqrt{\rho_1\rho_2\rho_2}}{\rho}\sqrt{(\cos\theta-\cos\theta_1)(\cos\theta_2-\cos\theta)},\\
    \frac{d\rho}{d\xi}&=-2\sqrt{(\rho-\rho_1)(\rho_2-\rho)(\rho_3-\rho)}
\end{split}
\end{equation}
identical to the equation obtained for this special case in Refs.~\cite{wright-13,kamch-13,kamch-14,kamch-14b}.

It is worth noticing that the integrals of motion (\ref{eq10}) and (\ref{eq11}) can be
cast after introduction of these new variables $\rho$ and $\theta$ to the form
\begin{equation}\label{eq13}
\begin{split}
    &H=\frac{\rho_{\xi}^2}{8\rho}-\frac{\rho^2}2+\tilde{\be}\rho\\
    &+
    \frac{\rho^2\sin^2\theta\cdot\theta_{\xi}^2+8[\al_1^2+\al_2^2-(\al_1^2-\al_2^2)\cos\theta]}
    {8\rho\sin^2\theta}=h,
    \end{split}
\end{equation}
\begin{equation}\label{eq14}
\begin{split}
    K=&\frac{\rho^2\sin^2\theta\cdot\theta_{\xi}^2+8[\al_1^2+\al_2^2-(\al_1^2-\al_2^2)\cos\theta]}
    {4\sin^2\theta}\\
    &-(\al_1+\al_2)^2=k.
    \end{split}
\end{equation}
The angle $\theta$ can be excluded from Eq.~(\ref{eq13}) with the use of Eq.~(\ref{eq14}) what gives
the equation for a single variable $\rho$,
\begin{equation}\label{eq15}
    \rho_{\xi}^2=4\rho^3-8\tilde{\be}\rho^2+8h\rho-4[k+(\al_1+\al_2)^2]
\end{equation}
which is another form of the second equation (\ref{eq-l4}).
Its solution can be expressed in terms of elliptic functions. When $\rho$ is known, then $\theta=\theta(\xi)$
can be obtained by integration of the equation (\ref{eq14}) or
\begin{equation}\label{eq16}
    \theta_{\xi}^2=\frac8{\rho^2}\left[\frac{k+(\al_1+\al_2)^2}2-\frac{\al_1^2+\al_2^2-(\al_1^2-\al_2^2)\cos\theta}
    {\sin^2\theta}\right],
\end{equation}
which also coincides up to the notation with the first equation (\ref{eq-l4}).
When their solutions are found then the components densities are given by (\ref{eq-l2}) transformed to
\begin{equation}\label{eq12}
    \rho_1(\xi)=\rho(\xi)\cos^2\frac{\theta(\xi)}2,\quad
    \rho_2(\xi)=\rho(\xi)\sin^2\frac{\theta(\xi)}2
\end{equation}
More details about these solutions can be found in Refs.~\cite{wright-13,kamch-13,kamch-14,kamch-14b}.
In particular, they include also the well-known Manakov dark-dark soliton solution.

\subsection{Appelrot-Delone class of solutions}

The systems (\ref{eq26}) and (\ref{eq26a}) were applied for the first time to a real mechanical problem by Sophie Kowalevski
in her theory of rotation of the so-called Kowalevski top \cite{kow-1889}. After that some particular
especially remarkable motions of this top were discussed by other authors, in particular,
by G.~G.~Appelrot and N.~B.~Delone (see, e.g., \cite{golubev}). Here we shall apply their method to the special case of
nonlinear motion of a two-component BEC which we shall also call the {\it Appelrot-Delone case}.

Let us suppose that the polynomial $\mathcal{R}(q)$ has a double root $q=\bar{\nu}$, the other roots we denote
$\nu_1\leq\nu_2\leq\nu_3$, that is we have
\begin{equation}\label{eq28}
    \mathcal{R}(q)=(q-\bar{\nu})^2(q-\nu_1)(q-\nu_2)(q-\nu_3)\equiv(q-\bar{\nu})^2\mathcal{R}_1(q),
\end{equation}
where $\mathcal{R}_1(q)$ is the 3rd degree polynomial with the roots $\nu_1,\,\nu_2,\,\nu_3$.
In this case it is convenient to use the system (\ref{eq26a}) written now in the form
\begin{equation}\label{eq29}
    \begin{split}
    2(q_1-\bar{\nu})\sqrt{\mathcal{R}_1(q_1)}&=(q_1-q_2)\frac{dq_1}{d\xi},\\
    2(q_2-\bar{\nu})\sqrt{\mathcal{R}_1(q_2)}&=-(q_1-q_2)\frac{dq_2}{d\xi}.
    \end{split}
\end{equation}
It is easy to see that this system is satisfied if
\begin{equation}\label{eq30}
    \begin{split}
    &q_1=\bar{\nu},\quad \frac{dq_2}{d\xi}=2\sqrt{\mathcal{R}_1(q_2)}\\
    \text{or}\quad &q_2=\bar{\nu},\quad \frac{dq_1}{d\xi}=2\sqrt{\mathcal{R}_1(q_1)}.
    \end{split}
\end{equation}
Both solutions lead to the same physical solution due to symmetry of Eqs.~(\ref{eq17})
with respect to transposition of $q_1$ and $q_2$. For definiteness we shall take the
second solution in (\ref{eq30}). Then the variable $q_1$ oscillates in the interval
$\nu_1\leq q_1\leq\nu_2$, where $\mathcal{R}_1(q_1)\geq0$ and, hence, for $\be>0$, we have
according to (\ref{eq27}) two choices for the parameters $\bar{\nu},\,\nu_1,\,\nu_2$,
\begin{equation}\label{eq31}
\begin{split}
    &\be\leq\nu_1<\nu_2,\quad-\be\leq\bar{\nu}\leq\be\\
    -&\be\leq\nu_1<\nu_2\leq\be,\quad\bar{\nu}\geq\be.
    \end{split}
\end{equation}
As we shall see, the second choice cannot give the soliton solution, so we shall consider
the first one.

In standard way we obtain
\begin{equation}\label{eq32}
    q_1(\xi)=\nu_1+(\nu_2-\nu_1)\mathrm{sn}^2(\sqrt{\nu_3-\nu_1}\,(\xi-\xi_0),m),
\end{equation}
where
\begin{equation}\label{eq33}
    m=\frac{\nu_2-\nu_1}{\nu_3-\nu_1},
\end{equation}
and, to simplify the notation, from now on we shall put the integration constant $\xi_0=0$.
Then the components densities are given by
\begin{equation}\label{eq34}
    \begin{split}
    &\rho_1=\frac{\bar{\nu}+\be}{2\be}[\be+\nu_1+(\nu_2-\nu_1)\mathrm{sn}^2(\sqrt{\nu_3-\nu_1}\,\xi,m)],\\
    &\rho_2=\frac{\bar{\nu}-\be}{2\be}[\be-\nu_1-(\nu_2-\nu_1)\mathrm{sn}^2(\sqrt{\nu_3-\nu_1}\,\xi,m)],
    \end{split}
\end{equation}
and their substitution into (\ref{eq6a}) yields the flow velocities. These formulae represent
the periodic nonlinear wave which can be called the {\it density wave}, since the densities
oscillate in phase and in the small amplitude limit this wave reduces to the sound wave, which
describes oscillations of the total density $\rho=\rho_1+\rho_2$.

\begin{figure}[ht]
\begin{center}
\includegraphics[width=6cm]{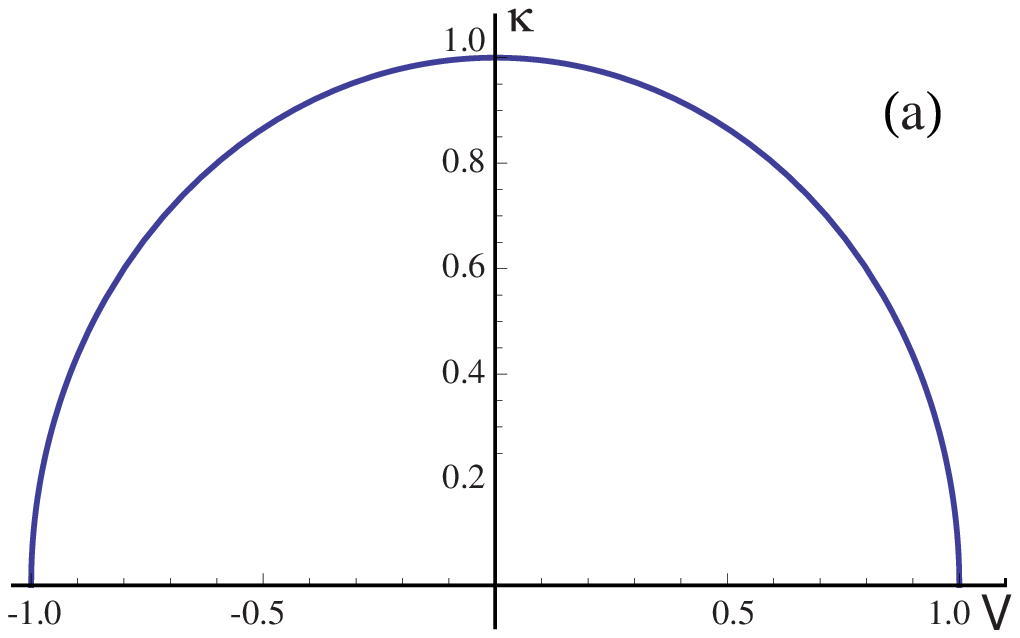}\\
\includegraphics[width=6cm]{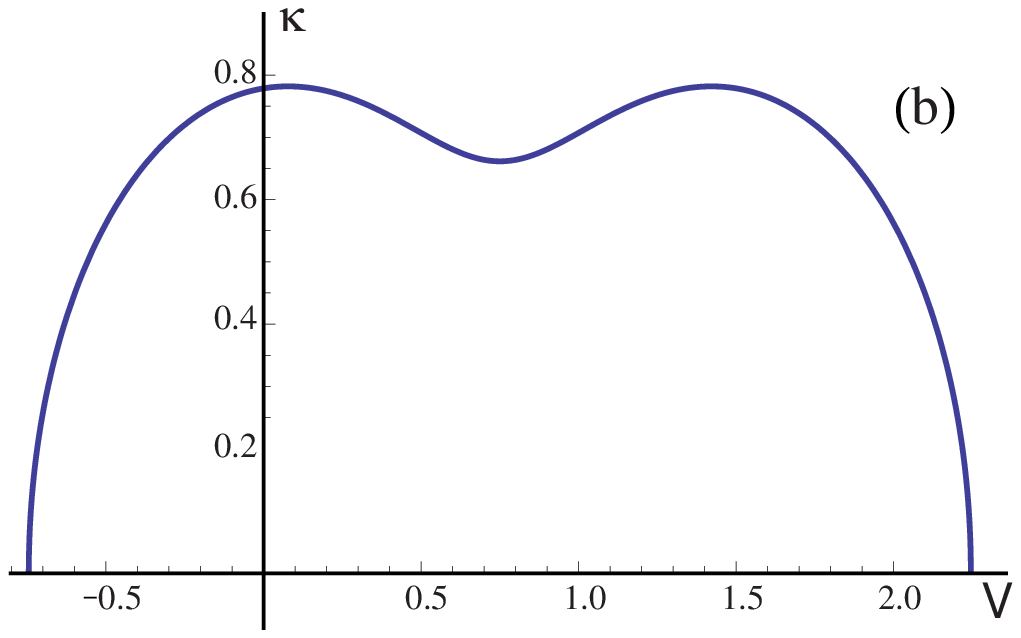}\\
\includegraphics[width=6cm]{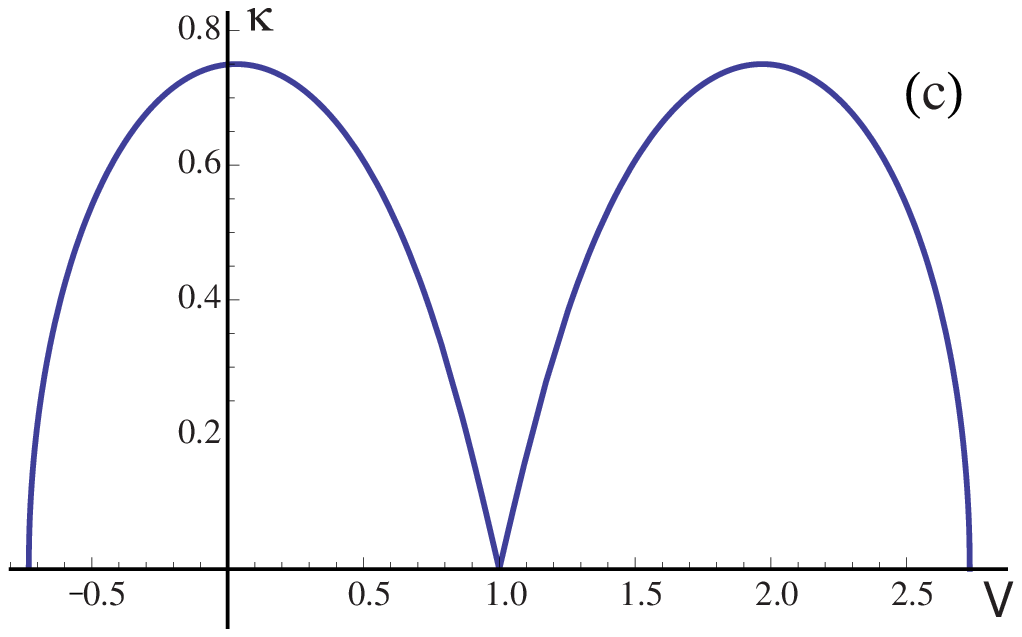}\\
\includegraphics[width=6cm]{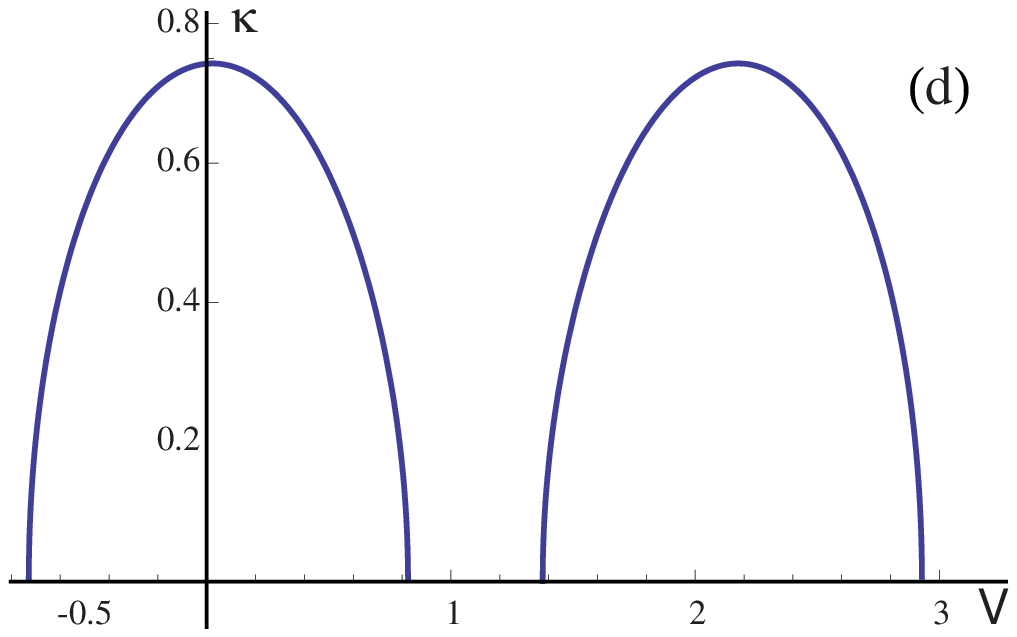}
\caption{Dependence of the inverse width $\kappa$ of a dark-dark soliton on its velocity $V$ for different values of the
relative velocity between condensates. In all plots $\rho_{10}=\rho_{20}=0.5$, $u_{10}=0$ and $V$ is measured with respect to
quiescent condensate:
(a) $u_{20}=0$, this is the case of well-known dark-dark solitons propagating through two still condensates;
(b) plot corresponding to $u_{20}=1.5$ illustrates a non-monotonic dependence;
(c) at $u_{20}=2.0$ the region of possible velocities splits with formation of two disconnected regions;
(d) plot for $u_{20}=2.2$ illustrates appearance of two disconnected regions of possible soliton velocities $V$.
}
\end{center}
\label{fig2}
\end{figure}

Let us consider the soliton limit when $\nu_3\to\nu_2$ ($m\to1$):
\begin{equation}\label{eq35}
    \begin{split}
    &\rho_1=\frac{\bar{\nu}+\be}{2\be}\left(\be+\nu_2-\frac{\nu_2-\nu_1}{\cosh^2(\sqrt{\nu_2-\nu_1}\xi)}\right),\\
    &\rho_2=\frac{\bar{\nu}-\be}{2\be}\left(\be-\nu_2+\frac{\nu_2-\nu_1}{\cosh^2(\sqrt{\nu_2-\nu_1}\xi)}\right),
    \end{split}
\end{equation}
where $\be<\nu_1<\nu_2$ and $-\be<\bar{\nu}<\be$. These parameters can be expressed in terms of the constant
densities at $|\xi|\to\infty$,
\begin{equation}\label{eq36}
    \rho_{10}=\frac1{2\be}(\bar{\nu}+\be)(\be+\nu_2),\quad \rho_{20}=\frac1{2\be}(\bar{\nu}-\be)(\be-\nu_2).
\end{equation}
Solving this system with respect to $\nu_2$ and $\bar{\nu}$ gives
\begin{equation}\label{eq37}
    \begin{split}
    \nu_2&=\frac12\left(\rho_{10}+\rho_{20}-\sqrt{(\rho_{10}-\rho_{20}-2\be)^2+4\rho_{10}\rho_{20}}\right),\\
    \bar{\nu}&=\frac12\left(\rho_{10}+\rho_{20}+\sqrt{(\rho_{10}-\rho_{20}-2\be)^2+4\rho_{10}\rho_{20}}\right).
    \end{split}
\end{equation}
The parameter $\be$ can be also expressed in terms of $\rho_{10},\,\rho_{20}$ and constant flow velocities
$u_{10},\,u_{20}$ at $|\xi|\to\infty$. From (\ref{eq6a}) and (\ref{eq6b}) we get
\begin{equation}\label{eq38}
    \al_1=\rho_{10}(u_{10}-V),\quad \al_2=\rho_{20}(u_{20}-V)
\end{equation}
and
\begin{equation}\label{eq39}
\begin{split}
    \be_1&=\frac12(u_{10}-V)^2+\rho_{10}+\rho_{20},\\
    \be_2&=\frac12(u_{20}-V)^2+\rho_{10}+\rho_{20},
    \end{split}
\end{equation}
hence
\begin{equation}\label{eq40}
    \be=\frac12\left[(u_{10}-V)^2-(u_{20}-V)^2\right].
\end{equation}
To determine the last unknown parameter $\nu_1$, we remark that Vi\`ete formula for the polynomial (\ref{eq25})
in our case gives $\be_1+\be_2=\nu_1+2(\nu_2+\bar{\nu})$ and, consequently,
\begin{equation}\label{eq41}
    \nu_1=\frac12\left[(u_{10}-V)^2+(u_{20}-V)^2\right].
\end{equation}
The solution (\ref{eq35}) exists if $\nu_2>\nu_1$ and this condition gives restrictions for the
soliton velocity,
\begin{equation}\label{eq42}
    \frac{\rho_{10}}{(V-u_{10})^2}+\frac{\rho_{20}}{(V-u_{20})^2}>1.
\end{equation}
Note that for the second choice in (\ref{eq31}) the condition $\nu_2>\nu_1$ cannot be fulfilled.
In the limiting case of one-component quiescent condensate ($\rho_{20}=0,\,u_{10}=0$) the condition
(\ref{eq42}) reduces to
the well-known fact that the soliton velocity is smaller than the sound velocity, $V<c_s=\sqrt{\rho_{10}}$.

Dependence of the inverse width $\kappa=\sqrt{\nu_2-\nu_1}$ of soliton on its velocity $V$ is given by
\begin{equation}\label{eq43}
\begin{split}
    &\kappa=\frac1{\sqrt{2}}\Big\{\rho_{10}+\rho_{20}-(u_{10}-V)^2-(u_{20}-V)^2\\
    &+\sqrt{[\rho_{10}-\rho_{20}-(u_{10}-V)^2+(u_{20}-V)^2]^2+4\rho_{10}\rho_{20}}\Big\}^{1/2}.
    \end{split}
\end{equation}
This expression can be also obtained by linearization of equations (\ref{eq6b}) with respect to
small deviations $\rho_k^{\prime}$ around asymptotic densities ($\rho_k=\rho_{k0}+\rho_k^{\prime}$)
and seeking the solution of the linearized equations in the form $\rho_k^{\prime}\propto \exp(-\kappa|\xi|)$.
This calculation shows that the Appelrot-Delone class of solutions yields in the corresponding limit
{\it all} soliton solutions with exponentially decaying tails around {\it non-zero} background
densities $\rho_{k0}\neq 0$.

Dependence (\ref{eq43}) is illustrated in Fig.~2 for different values of the relative velocity
of the BEC components. The remarkable new feature is that this dependence can be non-monotonic and
for large enough values of the relative velocity the region of possible values of the velocity $V$
splits into two separated regions in sharp contrast with the one-component situation. The appearance of
two regions of velocity can be illustrated graphically in the following way. We introduce for convenience
the variables $X=V-u_{10}$, $Y=V-u_{20}$; then the boundary of the region (\ref{eq42}) is given
by the equation
\begin{equation}\label{eqC1}
    X^2Y^2-\rho_{10}Y^2-\rho_{20}X^2=0.
\end{equation}
Its plot is shown in Fig.~3 by a solid line and the admissible values of $V$ are located inside this
line (that is in the area including the origin of the coordinate system). If we fix the value of
the relative velocity $U_0=u_{20}-u_{10}\equiv X-Y$, then the possible values of $V$ correspond
to points of the straight line located between its intersections with the curve (\ref{eqC1}).
Consequently, the splitting of the region of possible values of $V$ correspond to such $U_0$
that the straight line $X-Y=U_0$  touches the curve (\ref{eqC1})
at the point where $dY/dX=1$ or
\begin{equation}\label{eqC2}
    XY^2+X^2Y-\rho_{10}Y-\rho_{20}X=0
\end{equation}
(see Fig.~3).
The system (\ref{eqC1}) and (\ref{eqC2}) can be easily solved to give
\begin{equation}\label{eqC3}
\begin{split}
    &X=V-u_{10}=\pm\rho_{10}^{1/3}\sqrt{\rho_{10}^{1/3}+\rho_{20}^{1/3}},\\
    &Y=V-u_{20}=\mp\rho_{20}^{1/3}\sqrt{\rho_{10}^{1/3}+\rho_{20}^{1/3}},
    \end{split}
\end{equation}
and hence the critical value of the relative velocity is given by
\begin{equation}\label{eqC4}
    U_0=(\rho_{10}^{1/3}+\rho_{20}^{1/3})^{3/2}.
\end{equation}

\begin{figure}[ht]
\begin{center}
\includegraphics[width=6.5cm]{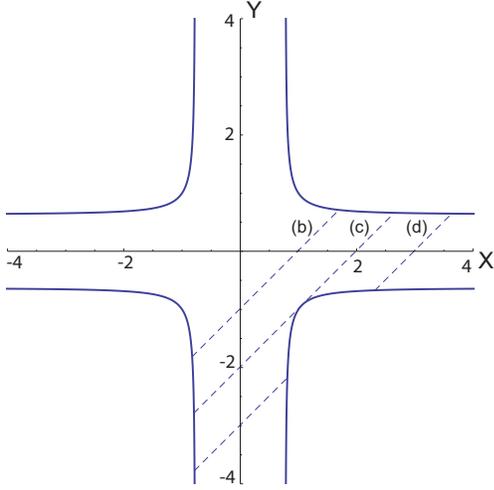}\\
\caption{Plot of the curve (\ref{eq43a}) (solid line) and of the straight lines $X-Y=U_0$
(dashed lines): (b) corresponds to a single region of possible values of
$V$ (see Fig.~2b); (c) correspond to the relative velocity at which the region splits
into two regions (see Fig.~2c); (d) corresponds to two separated regions (see Fig.~2d).
}
\end{center}
\label{fig3}
\end{figure}

\subsection{Dark-bright soliton solution}

If one of the background densities vanishes (say, $\rho_{20}=0$), then
the so-called dark-bright soliton solutions of the Manakov system are obtained (see, e.g.,
\cite{sk-1997,{smkj-10}}). Here we show that this type of solutions is a specialization of
general solutions of the Kowalevski equations when the polynomial $\mathcal{R}(q)$ has two
double zeroes.
In this case the condition (\ref{eq27b}) is fulfilled if one of the double
zeroes coincides with $\be$. Thus, we assume that
\begin{equation}\label{db1}
    -\be\leq\nu_1\leq q_1\leq\be\quad\text{and}\quad \be\leq q_2\leq\bar{\nu}=\nu_4=\nu_5.
\end{equation}
Then the system (\ref{eq26a}) reduces to
\begin{equation}\label{db2}
    \begin{split}
    2(\be-q_1)(\bar{\nu}-q_1)\sqrt{q_1-\nu_1}&=(q_1-q_2)\frac{dq_1}{d\xi},\\
    2(q_2-\be)(\bar{\nu}-q_2)\sqrt{q_2-\nu_1}&=-(q_1-q_2)\frac{dq_2}{d\xi}.
    \end{split}
\end{equation}
As in the preceding subsection, we see that the second equation is satisfied identically by
$q_2=\bar{\nu}$ and the first equation $dq_1/d\xi=-2(\be-q_1)\sqrt{q_1-\nu_1}$ can be easily
integrated to give
$
q_1=\be-(\be-\nu_1)\cosh^{-2}(\sqrt{\be-\nu_1}\,\xi).
$
As a result we obtain the densities
\begin{equation}\label{db3}
    \begin{split}
    \rho_1&=(\bar{\nu}+\be)\left(1-\frac{(\be-\nu_1)/(2\be)}{\cosh^2(\sqrt{\be-\nu_1}\,\xi)}\right),\\
    \rho_2&=(\bar{\nu}-\be)\cdot\frac{(\be-\nu_1)/(2\be)}{\cosh^2(\sqrt{\be-\nu_1}\,\xi)}
    \end{split}
\end{equation}
which obviously correspond to the dark-bright soliton: the density $\rho_1$ has a dip at $\xi=0$
and approaches to the background density $\rho_0=\bar{\nu}+\be$ as $|\xi|\to\infty$
whereas $\rho_2$ has a hump at $\xi=0$ and vanishes as $|\xi|\to\infty$. Let us relate the
parameters of formulae (\ref{db3}) with standard physical parameters for the soliton solution.
To this end we define the inverse half-width $\kappa$ of the soliton by the equation $\kappa=\sqrt{\be-\nu_1}$
and introduce the ratio of the components densities at the center of the soliton
$\ga=(\rho_0-\rho_1(0))/\rho_2(0)=(\bar{\nu}+\be)/(\bar{\nu}-\be)$. Besides that we assume that there is
no flow of the first component at infinity: $u_{10}=0$. Then from Eqs.~(\ref{eq6a}) and (\ref{eq6b}) we find
$\al_1=-V\rho_0$, $\al_2=0$, $\be_1=\rho_0+V^2/4$, $\be_2=\rho_0-\kappa^2/2$ and hence
\begin{equation}\label{db4}
    \be=\frac{V^2}4+\frac{\kappa^2}2=\rho_0\frac{\ga-1}{2\ga},\quad \nu_1=\be-\kappa^2,\quad \bar{\nu}=\rho_0\frac{\ga+1}{2\ga}.
\end{equation}
The dependence of the soliton's inverse width on its velocity is given by the formula
\begin{equation}\label{db5}
    \kappa(V)=\sqrt{\frac{\ga-1}{\ga}\rho_0-\frac{V^2}2}.
\end{equation}
These formulae are equivalent to those found in Ref.~\cite{sk-1997}.

\subsection{Legendre-Jacobi class of solutions}

The general one-phase traveling waves described by the Manakov system can be illustrated by an
easy numerical solutions of the Kowalevski equations (\ref{eq26a}). On the other hand, the analytical
solution (\ref{eq27a}) can be expressed in terms of Riemann $\theta$-functions by the methods
used already by S.~Kowalevski (see \cite{kow-1889,golubev}) and developed further in the algebraic-geometric
approach to integrable equations (see, e.g., \cite{bbeim}). This method was applied to the one-phase solutions of
the focusing Manakov system in Refs.~\cite{cezk-2000,eek-2000}.
However, the resulting
expressions are quite inconvenient for practical use. Therefore we shall confine ourselves here to
a particular case, when the solution can be reduced to the much better known special functions
(elliptic integrals) which permit one to understand the characteristic features of the solution
in a much simpler way. Here we shall consider such a
situation first noticed by Legendre \cite{legendre} and generalized by Jacobi \cite{jacobi-1832}.

Let the zeroes of the polynomial $\mathcal{R}(q)$ be given by
\begin{equation}\label{eq43a}
\begin{split}
    &\nu_1=c-1/b,\quad \nu_2=c-1/a,\quad \nu_3=c,\\
    &\nu_4=c+1,\quad \nu_5=c+1/ab,
    \end{split}
\end{equation}
where $0<b\leq a \leq 1$ and the parameter $c$ satisfies the conditions
\begin{equation}\label{eq43b}
    \mathrm{max}\{\beta,\,1/b-\beta\}< c< \beta+1/a.
\end{equation}
so that $q_1$ and $q_2$ oscillate within the intervals
\begin{equation}\label{eq43c}
    -\beta<\nu_1\leq q_1\leq \nu_2<\beta,\quad
    \beta<\nu_3\leq q_2\leq \nu_4.
\end{equation}
Let us assume for definiteness that at $\xi=0$ we have $q_1(0)=c-1/a$ and $q_2=c$ (other choices of the initial
conditions can be considered in a similar way).
Then, introducing the variables
\begin{equation}\label{eq43d}
z_{1,2}=q_{1,2}-c,
\end{equation}
we represent the solution (\ref{eq27a}) in the form
\begin{equation}\label{eq44}
\begin{split}
    &\int_{-1/a}^{z_1}\frac{dz}{\sqrt{\widetilde{\mathcal{R}}(z)}}+\int_{0}^{z_2}\frac{dz}{\sqrt{\widetilde{\mathcal{R}}(z)}}=0,\\
    &\int_{-1/a}^{z_1}\frac{zdz}{\sqrt{\widetilde{\mathcal{R}}(z)}}+\int_{0}^{z_2}\frac{zdz}{\sqrt{\widetilde{\mathcal{R}}(z)}}=\pm2{\xi},
     \end{split}
\end{equation}
where
\begin{equation}\label{eq45}
    \widetilde{\mathcal{R}}(z)=z(1-z)(1-abz)((1+az)(1+bz)/(ab)^2.
\end{equation}
As Jacobi showed \cite{jacobi-1832}, the integrals here can be calculated in terms of incomplete elliptic
integrals of the first kind. Since Jacobi did not provide the details of his method, this calculation is discussed
briefly in Appendix. As a result, we obtain a particular solution of Eqs.~(\ref{eq26}) in the form
\begin{equation}\label{eq46}
    \begin{split}
   &F(\vphi_{1a},k_1)+F(\vphi_{1b},k_2)-F(\vphi_2,k_1)-F(\vphi_2,k_2)=0,\\
   &F(\vphi_{1a},k_1)-F(\vphi_{1b},k_2)-F(\vphi_2,k_1)+F(\vphi_2,k_2)\\
   &=\pm 2 \sqrt{ \frac{(1+a)(1+b)}{ab} }\,\xi,
    \end{split}
\end{equation}
where
\begin{equation}\label{eq47}
    \vphi_{1a}=\left\{
    \begin{array}{l}
    \pi-\arcsin\sqrt{\frac{(1+az_1)(1+bz_1)}{z_1(\sqrt{a}-\sqrt{b})^2}},\quad -\frac1b\leq z_1\leq -\frac1{\sqrt{ab}},\\
    \arcsin\sqrt{\frac{(1+az_1)(1+bz_1)}{z_1(\sqrt{a}-\sqrt{b})^2}},\quad -\frac1{\sqrt{ab}}\leq z_1\leq-\frac1a;
    \end{array}
    \right.
\end{equation}
\begin{equation}\label{eq48}
    \vphi_{1b}=\arcsin\sqrt{\frac{(1+az_1)(1+bz_1)}{z_1(\sqrt{a}+\sqrt{b})^2}},\quad -\frac1b \leq z_1\leq-\frac1a;
\end{equation}
\begin{equation}\label{eq49}
    \vphi_{2}=\arcsin\sqrt{\frac{(1+a)(1+b)z_2}{(1+az_2)(1+bz_2)}},\quad 0 \leq z_2\leq 1.
\end{equation}
These equations determine implicitly the dependence of $z_1$ and $z_2$, and, hence, of $q_1$ and $q_2$, on $\xi$
in the interval of $\xi$ until the first turning point is met ($z_1=-1/b$ or $z_2=1$). After that the sign before the
corresponding square root in the Kowalevski equations (\ref{eq26a}) must be changed and the replacement
in the solution (\ref{eq44})
$$
    \int_{-1/a}^{z_1}\frac{dz}{\sqrt{\widetilde{\mathcal{R}}(z)}}\to \int_{-1/a}^{-1/b}\frac{dz}{\sqrt{\widetilde{\mathcal{R}}(z)}}-
\int_{-1/b}^{z_1}\frac{dz}{\sqrt{\widetilde{\mathcal{R}}(z)}}
$$
or
$$
\int_{0}^{z_2}\frac{dz}{\sqrt{\widetilde{\mathcal{R}}(z)}}\to \int_{0}^{1}\frac{dz}{\sqrt{\widetilde{\mathcal{R}}(z)}}-
\int_{1}^{z_2}\frac{dz}{\sqrt{\widetilde{\mathcal{R}}(z)}}
$$
must be done with similar changes in the expressions (\ref{eq46}). Making such changes at every successive turning
point, we find the solution in any necessary interval
of $\xi$. Substitution of resulting $q_1=z_1+c$ and $q_2=z_2+c$ into Eqs.~(\ref{eq17}) yields the dependence of
densities $\rho_1$ and $\rho_2$ on $\xi$. Typical resulting plots are shown in Fig.~4.

\begin{figure}[ht]
\begin{center}
\includegraphics[width=6.5cm]{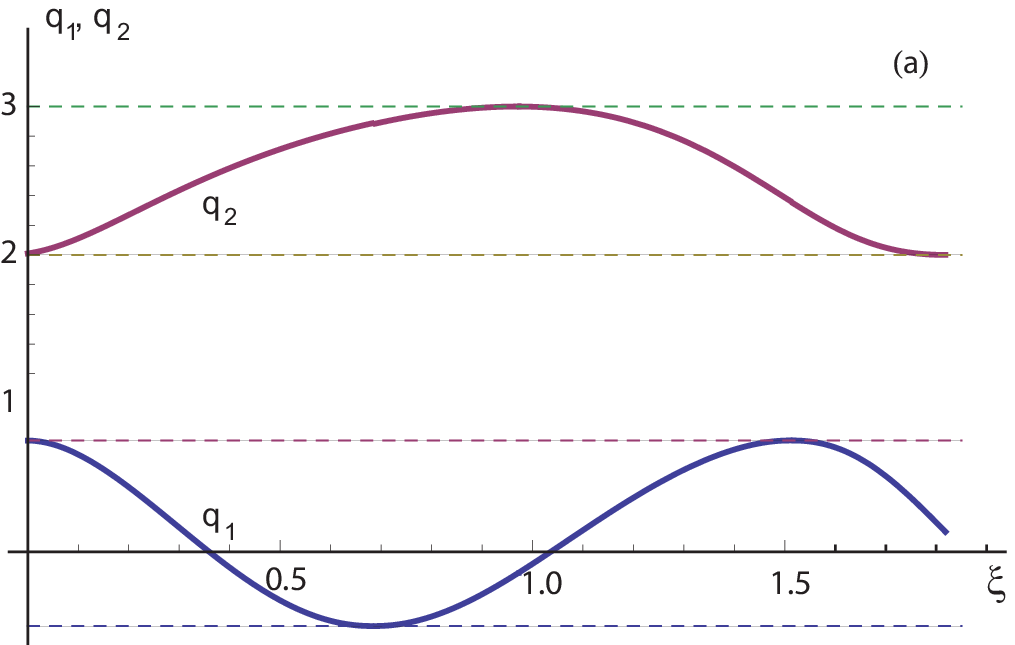}\\
\vspace{0.5cm}
\includegraphics[width=6.5cm]{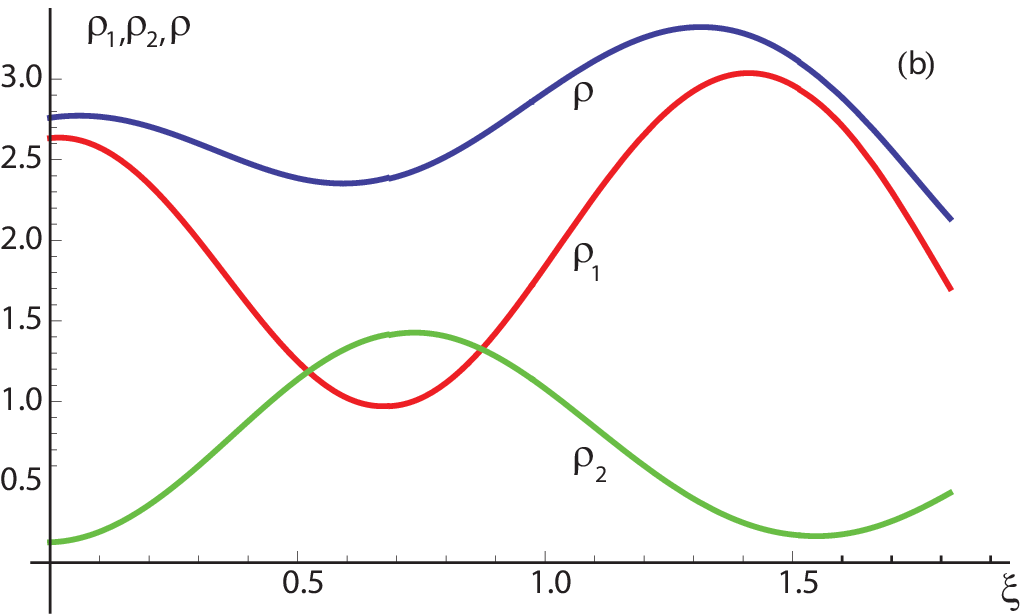}
\caption{(Color online) Plots for the solution (\ref{eq46}) of the Kowalevski equations for the values of
the parameters $a=0.8$, $b=0.4$, $c=2$
which correspond to $\nu_1=-0.5$, $\nu_2=0.75$, $\nu_3=2.0$, $\nu_3=3.0$, $\nu_5=5.125$. The initial conditions
are given by $q_1(0)=\la_2$, $q_2(0)=\la_3$.
(a) Plots of $q_1$ and $q_2$ for the interval of $\xi$ corresponding to full cycle of $q_2$-variable;
dashed lines indicate the values of $\nu_i$, $i=1,2,3,4$;
(b) plots of the components densities $\rho_1$ and $\rho_2$ and of the total density $\rho=\rho_1+\rho_2$.
}
\end{center}
\label{fig4}
\end{figure}

As one can see in Fig.~4, in the general solution the periodicity of the wave in space and time is lost
and the wave pattern demonstrates quite complicated behavior as a function of $\xi$.

\section{Conclusion}

In this paper we have found the one-phase traveling wave solution of the Manakov system which describes evolution of
two-component BEC. It is shown that in this case the Manakov system reduces to the equations which S.~Kowalevski
derived in her study of rotation of a heavy top in the discovered by her completely integrable case.
We show that the previously found solutions of the Manakov system  appear in this scheme as particular cases.
Besides that, new solutions are found which were either missed in previous analysis or cannot be obtained
by more elementary methods when parameters of the solutions are chosen in such a way that the evolution
equations are greatly simplified.
In particular, we have found a new dark-dark soliton solution for a two-component
BEC with the non-zero relative motion of the components. This solution has very unusual dependence of
the inverse width on the soliton's velocity. In principle, this can lead to new forms of dispersive
shock waves evolved from initial step-like distributions of the components densities or velocities.

For applications of the developed theory to the description of the polarization wave patterns observed in the
experiments \cite{hamner-11,hoefer-11,hamner-13}, the Whitham modulation theory \cite{whitham,kamch-2000} for these waves
has to be developed. Some particular situations have already been studied in Refs.~\cite{wright-95}
(genus-zero case) and \cite{kamch-14b} (genus-one case for the limit $\be=0$). The results of the present paper
demonstrate that the general one-phase solution is described by the fifth-degree polynomial $\mathcal{R}(q)$ whose zeroes
as well as the wave velocity must be related in framework of the finite-gap integration method with the modulation
parameters appearing in the Whitham theory of modulations of nonlinear waves.
Thus, the results obtained here provide the necessary step to development of the
modulation theory which can be applied to description of dispersive polarization shock waves observed
experimentally. Derivation of the Whitham equations is a difficult problem far beyond the scope of this
paper and we hope to consider it elsewhere.

\begin{acknowledgments}
  We thank E.~V.~Ferapontov, M.~A.~Hoefer, M.~P.~Kharlamov and M.~V.~Pavlov  for useful
  discussions.
\end{acknowledgments}

\setcounter{equation}{0}

\renewcommand{\theequation}{A.\arabic{equation}}

\section*{Appendix}

We have to calculate the integrals
\begin{equation}\label{A1}
\begin{split}
    &I_1=\int_0^z\frac{dz}{\sqrt{\widetilde{\mathcal{R}}(z)}},\quad
    I_1^{\prime}=\int_0^z\frac{zdz}{\sqrt{\widetilde{\mathcal{R}}(z)}},\\
    &I_2=\int_{-1/a}^z\frac{dz}{\sqrt{\widetilde{\mathcal{R}}(z)}},\quad
    I_2^{\prime}=\int_{-1/a}^z\frac{zdz}{\sqrt{\widetilde{\mathcal{R}}(z)}}.
    \end{split}
\end{equation}
The integrals $I_1$ and $I_1^{\prime}$ are calculated with the use of the substitution
\begin{equation}\label{A2}
    1+abz^2=uz
\end{equation}
or
\begin{equation}\label{A3}
    \sqrt{z}=\left(\sqrt{u+2\sqrt{ab}}\pm\sqrt{u-2\sqrt{ab}}\right)/(2\sqrt{ab}),
\end{equation}
where $u$ is a new integration variable. It is easy to see that the function (\ref{A3})
with the lower sign maps the interval $1+ab\leq u<\infty$ on $0\leq z\leq 1$ and with
the upper sign maps the same interval on $1/(ab)\leq z<\infty$. Then substitution
(\ref{A3}) with the lower sign into $I_1$ gives after simple manipulations
\begin{equation}\label{A4}
    \begin{split}
    I_1&=\frac{ab}2\int_{u(z)}^{\infty}\frac{du}{\sqrt{(u+a+b)(u+2\sqrt{ab})(u-1-ab)}}\\
    &+\frac{ab}2\int_{u(z)}^{\infty}\frac{du}{\sqrt{(u+a+b)(u-2\sqrt{ab})(u-1-ab)}},
    \end{split}
\end{equation}
where
\begin{equation}\label{A5}
    u(z)=(1+abz^2)/z.
\end{equation}
Elliptic integrals in (\ref{A4}) are transformed to the standard form by the
substitution
\begin{equation}\label{A6}
    u=\frac{(1+a)(1+b)}{\sin^2\vphi}-a-b.
\end{equation}
As a result we obtain
\begin{equation}\label{A7}
    I_1=\frac{ab}{\sqrt{(1+a)(1+b)}}\left\{F(\vphi,k_1)+F(\vphi,k_2)\right\},
\end{equation}
where $F(\vphi,k)$ denotes the elliptic integral of the first kind,
\begin{equation}\label{A8}
    k_1=\frac{\sqrt{a}-\sqrt{b}}{\sqrt{(1+a)(1+b)}},\quad
    k_2=\frac{\sqrt{a}+\sqrt{b}}{\sqrt{(1+a)(1+b)}},
\end{equation}
and $\vphi$ is related with the upper limit of integration $z$ by the formula
\begin{equation}\label{A9}
    \sin^2\vphi=\frac{(1+a)(1+b)z}{(1+az)(1+bz)}.
\end{equation}
The integral $I_1^{\prime}$ is calculated by the same method and the result reads
\begin{equation}\label{A10}
    I_1^{\prime}=\sqrt{\frac{ab}{(1+a)(1+b)}}\left\{-F(\vphi,k_1)+F(\vphi,k_2)\right\}.
\end{equation}

The integrals $I_2$ and $I_2^{\prime}$ in (\ref{A1}) can be calculated with the use
of the substitution
\begin{equation}\label{A11}
    u(z)=-(1+abz^2)/z
\end{equation}
or
\begin{equation}\label{A12}
    \sqrt{-z}=\left(\sqrt{u+2\sqrt{ab}}\pm\sqrt{u-2\sqrt{ab}}\right)/(2\sqrt{ab}),
\end{equation}
which map the interval $2\sqrt{ab}\leq u\leq a+b$ on the intervals $-1/b\leq z\leq-1/\sqrt{ab}$
and $-1/\sqrt{ab}\leq z\leq-1/a$, correspondingly for upper and lower signs.
This transforms $I_2$ to
\begin{equation}\label{A13}
    \begin{split}
    I_2&=\frac{ab}2\int_{a+b}^{u(z)}\frac{du}{\sqrt{(u-a-b)(u+2\sqrt{ab})(u+1+ab)}}\\
    &+\frac{ab}2\int_{a+b}^{u(z)}\frac{du}{\sqrt{(u-a-b)(u-2\sqrt{ab})(u+1+ab)}}.
    \end{split}
\end{equation}
These integrals are reduced to standard form of elliptic integrals by substitutions
\begin{equation}\label{14}
    u=a+b-(a+b\pm2\sqrt{ab})\sin^2\vphi.
\end{equation}
As a result we obtain
\begin{equation}\label{A15}
    I_2=-\frac{ab}{\sqrt{(1+a)(1+b)}}\left\{F(\vphi_a,k_1)+F(\vphi_b,k_2)\right\},
\end{equation}
where
\begin{equation}\label{A16}
    \vphi_{a}=\left\{
    \begin{array}{l}
    \pi-\arcsin\sqrt{\frac{(1+az)(1+bz)}{z(\sqrt{a}-\sqrt{b})^2}},\quad -\frac1b\leq z\leq -\frac1{\sqrt{ab}},\\
    \arcsin\sqrt{\frac{(1+az)(1+bz)}{z(\sqrt{a}-\sqrt{b})^2}},\quad -\frac1{\sqrt{ab}}\leq z\leq-\frac1a;
    \end{array}
    \right.
\end{equation}
and
\begin{equation}\label{A17}
    \vphi_{b}=\arcsin\sqrt{\frac{(1+az)(1+bz)}{z(\sqrt{a}+\sqrt{b})^2}},\quad -\frac1b \leq z\leq-\frac1a.
\end{equation}
Similar calculation yields
\begin{equation}\label{A18}
    I_2^{\prime}=\sqrt{\frac{ab}{(1+a)(1+b)}}\left\{F(\vphi_a,k_1)-F(\vphi_b,k_2)\right\}.
\end{equation}

\end{document}